\documentclass[11pt,a4paper]{article}
\usepackage{my-own-commands}

\title{CMB anisotropies at all orders: the non-linear Sachs-Wolfe formula}

\author{Omar Roldan}

\affiliation{Instituto de F\'\i sica, Universidade Federal do Rio de Janeiro, 21941-972, \\ 
Rio de Janeiro, RJ, Brazil}

\emailAdd{oaroldan@if.ufrj.br}

\abstract{We obtain the non-linear generalization of the Sachs-Wolfe + integrated Sachs-Wolfe (ISW) formula describing the CMB temperature anisotropies. Our formula is valid at all orders in perturbation theory, is also valid in all gauges and includes scalar, vector and tensor modes. A direct consequence of our results is that the maps of the logarithmic temperature anisotropies are much cleaner than the usual CMB maps, because they automatically remove many secondary anisotropies. This can for instance, facilitate the search for primordial non-Gaussianity in future works. It also disentangles the non-linear ISW from other effects. Finally, we provide a method which can iteratively be used to obtain the lensing solution at the desired order.}


\begin{document}
\maketitle

\section{Introduction and main results}

The Cosmic Microwave Background (CMB) temperature anisotropies is one of the most important observables in cosmology. The CMB data is used for instance to constrain models of inflation \cite{INFLATION:Ade:2015lrj}, the amount of primordial non-Gaussianity \cite{NG:Ade:2015ava}, isocurvature perturbations \cite{INFLATION:Ade:2015lrj}, and to extract the cosmological parameters of the $\Lambda$CDM model \cite{PARAMETERS:Ade:2015xua}. Because of that, many efforts have been made to properly understand the physics of the CMB, this physics can be separated into three stages: before, during and after the period of recombination. Regarding the latter case, we can formally split the CMB anisotropies into  primary and secondary anisotropies. While the primary anisotropies (those already present at the emission time) are supposed to be known (for instance, by solving the Boltzmann and the Einstein's equations during and before recombination), secondary gravitational anisotropies must be obtained by solving the geodesic equation of photons in its way down to the observer. Note that additional anisotropies can arise due to secondary scatterings of photons with hot gas during the reheating period, however, this process is not covered here.

Temperature anisotropies were systematically investigated for the first time by Sachs and Wolfe \cite{Sachs:1967er} in 1967 by using first-order perturbation theory, and their famous formula is quite easy to understand
\begin{align}
	 \f{ \Delta T_o}{ \bar{T}_o} = \cT_e + (\Phi_e - \Phi_o) - ( v_e \cdot n_e  - v_o \cdot n_o) + I_0\,,
	\label{eq:SW}
\end{align}
where, the subscripts $ e$ and $ o$ means that quantities must be evaluated at the emission and observation event respectively. Here $ \cT, \Phi$ and $ I_0 $ are respectively the intrinsic temperature anisotropies, the gravitational potential and the integrated Sachs-Wolfe effect (ISW). The ISW is an integrated term due to the 
time variation of the metric perturbations along the path of the photon (it gives the accumulated redshift of photons when traveling along the evolving inhomogeneities). Finally, $ ( v_e \cdot n_e  - v_o \cdot n_o)$ gives the linear Doppler effect due to the observer's and emitter's peculiar velocity ($ v_o$ and $ v_e$), with $ n_o$ the direction of observation and $ - n_e$ the direction of emission.

The CMB temperature anisotropies are so small that the previous equation gives a very good description of the observed data, at least on large scales where secondary scatterings are negligible. Second-order perturbation theory of the CMB is however important to describe in a unified picture several important effects which are not taken into account by \eq{eq:SW}. These are for instance, lensing \cite{Seljak:1995ve,Zaldarriaga:1998ar}, time delay \cite{Hu:2001yq}, Doppler modulation and aberration \cite{Challinor:2002zh,Amendola:2010ty} and the Rees-Sciama effect \cite{Rees:1968zza,Crittenden:1995ak}. These effects although smaller than the first-order ones are very relevant for a correct understanding of the CMB physics, so second-order perturbation theory represents an essential tool for an accurate analysis of current and future CMB data. The full second-order generalization of the Sachs-Wolfe formula was obtained in 1997 by Mollerach and Matarrese \cite{Mollerach:1997up} by using a method introduced in \cite{Pyne:1993np,Pyne:1995bs}. Their second-order expression is somehow big and a direct interpretation of each term is difficult. Further progress in obtaining simple formulas have been given in \cite{Boubekeur:2009uk,Mirbabayi:2014hda}

In the search for non-Gaussianity, second-order perturbations is enough to study the three-point function (or its Fourier counterpart, the bispectrum). However, if one wants to go to the four-point function (or the trispectrum) for instance, third-order perturbation theory is needed to fully account for all the contributions. The CMB anisotropies up to third order were first computed in \cite{DAmico:2007ngk} by using two methods, the first one is the same used by \cite{Mollerach:1997up} in 1997, and the second one which is simpler and closer to our method, allowed them to obtain a fully non-linear Sachs-Wolfe formula for the specific case in which the metric is totally determined by two scalars variables, $ \Phi $ and $ \Psi$. Additional effort to obtain the non-linear description of the CMB can be found in \cite{Zibin:2008fe,Gao:2010ti,Saito:2014bxa}. 

\subsection{Main results: Discussion}

Choosing a particular parametrization of the metric is essential for obtaining exact solutions in cosmology, and this was the case in this work. By writing the line element as $ \dd s^2  = a^2(\eta) e^{2\Phi} \dd \hat{s}^2$ with the conformal metric given by
\begin{align}
	\dd \hat{s}^2  & \equiv - \dd \eta^2+ 2\beta_{j} \lp e^{-M} \rp_{ji} \ \dd x^i \dd \eta 
	+ \lp  e^{-2M}\rp_{ij}\dd x^i \dd x^j \,,
\end{align}
we were able to obtain an exact expression for the observed CMB temperature, $ T_o = \bar{T}_o\ e^{\Theta}$, where $ \bar{T}_o$ is the observed mean temperature and 
\begin{align}
	%
	\Theta & \equiv (\cT_e - \cT_o) + (\Phi_e - \Phi_o)  + I_0(x_{e},x_{o}) + 
	\ln \lp \f{\gamma_e \lp 1 - n_e \cdot v_e \rp}{\gamma_o \lp 1 - n_o \cdot v_o \rp} \rp 
	  \,.
	\label{eq:Gsw-intro}
\end{align}
Here $ \cT, \Phi$ and $ I_0$ are respectively, the non-linear generalization of the  intrinsic temperature anisotropies, the gravitational potential and the integrated Sachs-Wolfe effect. The logarithm term corresponds to the Doppler effect, with $ \gamma$ the Lorentz factor. $ x_e = ( \eta, x^{i})_e$ and $x_o = ( \eta, x^{i})_o $ are the spacetime coordinates of the emission and observation events, and $ \eta$ is the conformal time.

In terms of $ \Theta$, we can easily obtain the temperature anisotropies as
\begin{align}
	\f{ \Delta T_o}{ \bar{T}_o} = e^{\Theta} - 1 = \Theta + \f{\Theta^{2}}{2} + \cdots\,,
	\label{eq:DT2}
\end{align}
and in the case of perturbation theory we just need to truncate the series at the desired order. Because of the relation $ \Theta = \ln \lp 1 + \Delta T_o/ \bar{T}_o \rp $
\OLD{\begin{align}
	\Theta = \ln \lp \f{T_o}{\bar{T}_o} \rp  = \ln \lp 1 + \f{ \Delta T_o}{ \bar{T}_o} \rp \,,
\end{align}
}
we will call $ \Theta$ the \textit{logarithmic temperature anisotropies}. Note that:

\begin{itemize}
\item The intrinsic temperature anisotropies $ \cT_e$ are defined through the relation (this notation was also used in \cite{Gao:2010ti}) $ T_{e} = \braket{T}_{e} e^{ \cT_e }$ where $ \braket{T}_{e} $ is the background temperature at the time of emission. Note also the presence of the factor $ \cT_o$ in \eq{eq:Gsw-intro}, which is absent in previous works in literature. This factor is important for two reasons: it makes the expression for $ \Theta$ symmetric in the emission and observation points and it ensures the gauge invariance of $ \Theta$. Without this factor, neither $ \Theta$ nor $ \bar{T}_o$ would be gauge invariant (although the $ T_o$ would). So introducing $ \cT_o$ ensures also the gauge invariance of the mean temperature $ \bar{T}_o$ as it should be. The definition of $ \cT_o$ is given in the next section.

\item Although $ T_o$ contains crossed terms involving the fields at the emission and observation point (for instance, at second order it contains terms of the form $ \Phi_{e} \Phi_{o}$), $ \Theta$ does not contain such mixed terms. That is, $ \Theta$ is composed of a sum of locally defined terms. In particular the Doppler term is just: $ \ln \lp \gamma_o \lp 1 - n_o \cdot v_o \rp \rp - \ln \lp \gamma_e \lp 1 - n_e \cdot v_e \rp\rp$.

\item Note also that, in $ \Theta$ the ISW effect is clearly separated from other terms (although it is correlated with lensing \refs{sec:lensing}) like $ \Phi, \cT$ and $ v$. It makes the study of the ISW (as well as lensing) easier by directly using $ \Theta$ rather than $ \Delta T_o/ \bar{T}_o$. In previous expressions in literature (see for instance \cite{DAmico:2007ngk,Mollerach:1997up}), many integrated terms are coupled with other quantities making it difficult to isolate the ISW effect from the rest. So our results can be stated in a different way: by taking the logarithm of the temperature anisotropies we are making kind of ``resummations'' and removing these spurious non-linearities. This is similar to what happens in quantum field theory, in which the disconnected Feynman diagrams are removed by taking the logarithm of the propagators. Finally, since propagators in quantum mechanics are just correlation functions, we expect that the correlation functions of $ \Theta$ are much simpler than those of $ \Delta T_o/ \bar{T}_o$. For instance, by considering $ \Delta T_o/ \bar{T}_o$ instead of $ \Theta$, we are considering spurious quadratic, cubic, ..., terms which could create bias in the search for primordial non-Gaussianity.

\item Even if we treat $ \Theta$ as a first-order quantity, $ \Delta T_o/\bar{T}_o$ will not be  linear as it contains all powers of $ \Theta$. This shows that even if $ \Theta$ is a Gaussian distributed quantity (which in general is not the case, but it would be a good approximation if we evolve $ \Theta$ linearly from single-field initial conditions during inflation), $ \Delta T_o/\bar{T}_o$ is not Gaussian.%
\footnote{In particular, the off-diagonal part of the two-point correlation function does not vanish.
} %
Because of this, it seems better to use $ \Theta$ as the variable to be studied in future CMB experiments. That is, we propose to study the maps of the logarithmic temperature anisotropies $\ln \lp 1 + \Delta T_o/\bar{T}_o \rp $. Such a map will be free of Doppler modulation (see below) and other couplings%
\footnote{Note that the couplings induced by aberration and lensing cannot be removed by such procedure.
} %
which otherwise will be present in a normal map of $ \Delta T_o/\bar{T}_o$. In practice, what is measured in an experiment like Planck are the variations in the intensity
\begin{align}
	I_{obs} (\nu,n)= \f{2 h \nu^3 }{c^2 } { 1\over \exp \lp \f{h \nu}{k_B T_o(n)} \rp - 1}\,,
	\label{eq:Iobs}
\end{align}
so that $ \Theta$ can be calculated directly from the variation $ \delta I_{obs}$ without explicitly giving $ T_o$.

\item If we still want to analyze the data in terms of $ \Delta T/ \bar{T}$ rather than $ \Theta$, the theoretical n-point correlation function of $ \Delta T/ \bar{T}$ and $ \Theta$ are easily related for the specific case of a Gaussian distributed $ \Theta$ (see for instance \cite{Bartolo:2005fp}).

\item Within the linear regime, it is well known that for adiabatic perturbations, in the Poisson gauge and in the large scale limit (where we can neglect $ v_{e} \cdot n_{e}$ and the ISW term) we have $ \cT =  - 2 \Phi/3 $, so that $ \Theta \approx \Phi_{e}/3$ (here without considering the contributions at the observer). It has been shown in \cite{Bartolo:2005fp} that this relation continues to hold at the non-linear level. So, this result in conjunction with the formula \eq{eq:Gsw-intro} suggest that the metric parametrization introduced in this work and our definition of the non-linear intrinsic perturbations $ \cT$ are appropriated to extend the results of the linear theory to the non-perturbative level.

\item Since conformal transformations yield null geodesics into null geodesics, the photon's path is totally determined by the conformal metric $ \dd \hat{s}^2$. Therefore the integrated Sachs-Wolfe term $ I_0$, as well as the lensing terms encoded into $ x_e$ and $ n_{e}$ are totally determined by $ \beta_i$ and $ M_{ij}$. The explicit form of these quantities are given in the next sections.

\item In principle $ \Phi, \beta_i$ and $ M_{ij}$ are independent quantities, but in the linear regime (and during matter domination) general relativity predicts that $ M^{i}_{i}/3 = 2 \Phi$. So by measuring $ \Phi$ by an independent method like the use of the Poisson equation \cite{Amendola:2016saw}, and comparing with the lensing and ISW measurements, we can test general relativity. Note that by having the non-linear version of the ISW and lensing effects we could make a better interpretation of future data, this is because even if Einstein's gravity is correct, the inadequate use of the linear approximation to analyze the data could indicate a deviation from the expected relation between $ \Phi$ and $ M^{i}_{i}$.

\item An immediate consequence of \eq{eq:DT2} is that Doppler modulation of the temperature anisotropies always exist regardless of the nature of the dipole. Let's explain it a bit more. Split the logarithmic anisotropies as $ \Theta = \Theta_{d} + \tilde{\Theta}$, where in a multipolar expansion $ \Theta_{d}$ refers to the dipole of the logarithmic anisotropies and $ \tilde{\Theta}$ contains all the remaining multipolar components, that is, $ \ell \ge 2$. Then we see from \eq{eq:DT2} that the observed temperature anisotropies up to second order are given by
\begin{align}
	\f{ \Delta T_o}{ \bar{T}_o} = \Theta_{d} + \tilde{\Theta} + \f{ \tilde{\Theta}^2 + \Theta^2_d}{2} + \Theta_{d}\tilde{\Theta}\,.
	\label{eq:DT2v2}
\end{align}
The last term is what we call the Doppler modulation%
\footnote{The appropriate name will be dipolar modulation, but in the case in which the dipole is mainly of kinematical origin this modulation is due to the Doppler effect. We will adopt this name here because the CMB dipole is believed to be due to our peculiar velocity.
} %
 of the temperature anisotropies, and it leads to coupling between neighbors multipolar components ($ \ell, \ell \pm 1$) in the two-point correlation function that are proportional to the magnitude of the CMB dipole. These couplings (as well as aberration couplings) were measured by Planck%
\footnote{Two independent works realized that such effect could be observed by the Planck satellite, \cite{Kosowsky:2010jm,Amendola:2010ty}.
} %
in \cite{Aghanim:2013suk}. The results were consistent in amplitude and direction (at the $ 3 \sigma$-level) with the well known measured CMB dipole, that is, they are consistent with the prediction of simple formula $ \Theta_{d}\tilde{\Theta}$.

According to \eq{eq:DT2v2}, Planck's measurements tell us nothing about the nature of the CMB dipole. However, measuring Doppler modulation is important for the following reason: suppose that a more precise measurement of Doppler modulation is made by a future CMB experiment like CoRE \cite{Burigana:2017bxl}, suppose also that the results show a significant deviation from the simple expectation $ \Theta_{d}\tilde{\Theta}$, then that would imply that the term $ \tilde{ \Theta}$ in \eq{eq:DT2v2} contains dipolar-like modulation couplings, and they necessarily come from primordial non-Gaussianity terms that couple the long-mode (dipolar components) with the short-modes (the higher multipoles). Such a result will rule-out single-field-inflationary models and would require a non-negligible amplitude for the dipolar components. These facts were first noted in \cite{Roldan:2016ayx}. Although the previous results follows immediately from \eq{eq:DT2v2}, they were far from obvious by using previously existing formulas, like the one given in \cite{Mollerach:1997up}. Finally, we want to mention that the conclusions of \cite{Roldan:2016ayx} regarding dipolar modulation were restricted to the large scale case, but here the proof holds at any scale.

\end{itemize}

We want to stress that in our results we have assumed a perfect blackbody spectrum for the CMB (\eq{eq:Iobs}). However, it is known that spectral distortions (deviations from the blackbody spectrum) start been relevant at second order. A non-linear treatment of theses spectral distortions was introduced in the nice paper%
\footnote{I am very grateful to Cyril Pitrou for let me know about the works I cite in this paragraph.
} %
of Stebbins~\cite{Stebbins:2007ve}. There, Stebbins considered the observed spectrum as a superposition of blackbody with different temperatures and introduced the concept of \textit{mean logarithmic temperature} which must be related to our definition of $ \Theta$. On the other hand, in the same way as we are proposing the use $ \Theta = \ln \lp 1 + \Delta T_o/\bar{T}_o \rp $ for the future CMB maps, it was also advocated in \cite{Pitrou:2014ota} the use of the logarithmically averaged temperature moments to describe the spectral distortions. In \cite{Huang:2012ub} it was also noted the importance of the used of the exponential notation, though they considered particular cases.

Finally, note that in order to make quantitative predictions, the Sachs-Wolfe formula is not enough as we need to specify the fields $ \cT, \Phi, \beta^{i}, etc.$ as well as the integration path $ x^{i}( \eta)$. In this sense further progress is needed to obtain (analytical or numerical) non-linear solutions of the Einstein's (or Boltzmann's) equations. On the other hand, as perturbative solutions are still of high importance, in \refs{sec:SW2nd-curved} we obtain the full second-order Sachs-Wolfe formula. Perturbative solutions of the metric and fluid perturbations are known in some specific cases (e.g., by assuming matter domination or the large scale limit), see for instance \cite{Boubekeur:2008kn,Bartolo:2005kv}. For other useful results at second order, see \cite{Pitrou:2008hy,Senatore:2008vi,Bartolo:2006cu,Bartolo:2006fj}.

\subsection{Main results: Obtaining the generalized Sachs-Wolfe formula}
\label{sec:gSW}

In \refs{sec:geo-eq} we solve the time-component of the geodesic equation, which allow us to relate the observed temperature $ T_{o}$ with the emission temperature $ T_{e}$ by a simple relation. The results are given in \eq{eq:ToTeExpI0}, and can be expressed as%
\footnote{The explicit form of $ I_0$ is given in \eq{eq:I0}.
} %
\begin{align}
	T_{o} = T_{e} \f{a_e}{a_o}\ e^{\Phi_e - \Phi_o + I_{0}}\ 
	\f{\gamma_e \lp 1 - n_e \cdot v_e \rp}{\gamma_o \lp 1 - n_o \cdot v_o \rp}\,,
	\label{eq:ToTeExpI0-intro}
\end{align}
where $ v_{o}$ (and $ v_{e}$) is the \textit{peculiar velocity} of the observer (and the emitter). Note that, given an observer with four-velocity $ u$, its peculiar velocity is defined according to%
\footnote{Note that  $ v$ is related to $ \com{v}$, the velocity of  comoving-observers w.r.t $ u$ by the relation $ \com{v} = -v$, see \refs{sec:Photons-and-obs}.
} %
\begin{align}
	\com{u} = \gamma \lp u - v \rp \,, \qquad \gamma  = - 
	u \cdot \com{u} = \f{1}{ \sqrt{1 - v \cdot v }}\,,
	\label{eq:uv}
\end{align}
where, $ u \cdot v = 0$ and $ \com{u}$ is the four-velocity of comoving observers. In pp.5 of \cite{Stebbins:2007ve} a similar result to \eq{eq:ToTeExpI0-intro} was obtained but without including vector and tensor perturbations, and without considering the velocity of the emitter and observer.

We now define the \textit{logarithmic intrinsic temperature anisotropies} and clarify some concepts about the mean values. After that, we will get the final form of the generalized Sachs-Wolfe formula.

\subsection*{(Logarithmic) Intrinsic temperature anisotropies}

Before the epoch of recombination, the Universe was so hot and dense that photons frequently interacted with the free electrons via Thomson scattering,%
\footnote{See sections 8.7.1 and 11.3.1 of \cite{Lyth:2009zz} for further details on this.
} 
while the electrons frequently interacted with protons via Coulomb scattering, thus forming the so called photon-baryon fluid. As a result, the fluid reached a state of thermal equilibrium and the photons are well described by a blackbody \textit{distribution function}. However, because of the inhomogeneities the thermal equilibrium is just local, meaning that different local observers%
\footnote{Here, an observer can be for instance an electron.
} %
in the rest frame of the fluid will measure different values for the temperature $T$, that is, $T=T(x)$. During recombination the Compton scattering rate decreases and anisotropies in the photon's distribution function will appear, that is, $T=T(x,n)$.

We will write the temperature of the photon's fluid as
\begin{align}
	T( x, n) = \braket{ T}\ e^{ \cT} \,, 	\qquad	\textrm{where} \qquad \cT = \cT( x, n)\,,
\end{align}
and $ \braket{T} \propto 1/a(\eta)$ is the background temperature. We will call $ \cT$ the logarithm perturbations. This expression is meaningful for $ \eta \le \eta_e$, when the photons and baryons are still in equilibrium. That is, $ \cT$ is not defined for $ \eta > \eta_e$. Below we will provide an extension of $ \cT$ for $ \eta > \eta_e$, so that $ \cT$ is a field defined in the whole spacetime.

Note that $ \eta = const$ is defined by the physical argument that $ \braket{T} = const$ as in Mirbabayi \& Zaldarriaga \cite{Mirbabayi:2014hda}. In that sense, when transforming the CMB temperature, it is better to use gauge transformations (active transformations, acting on the fields) rather than passive transformations (transformations on the coordinates), because in the latter case the transformation of the time-coordinate becomes intricate. This issue will be discussed in details in a future work. See also \cite{Mirbabayi:2014hda} for an specific example.
\begin{figure}[h!]\centering
	\includegraphics[scale=0.35]{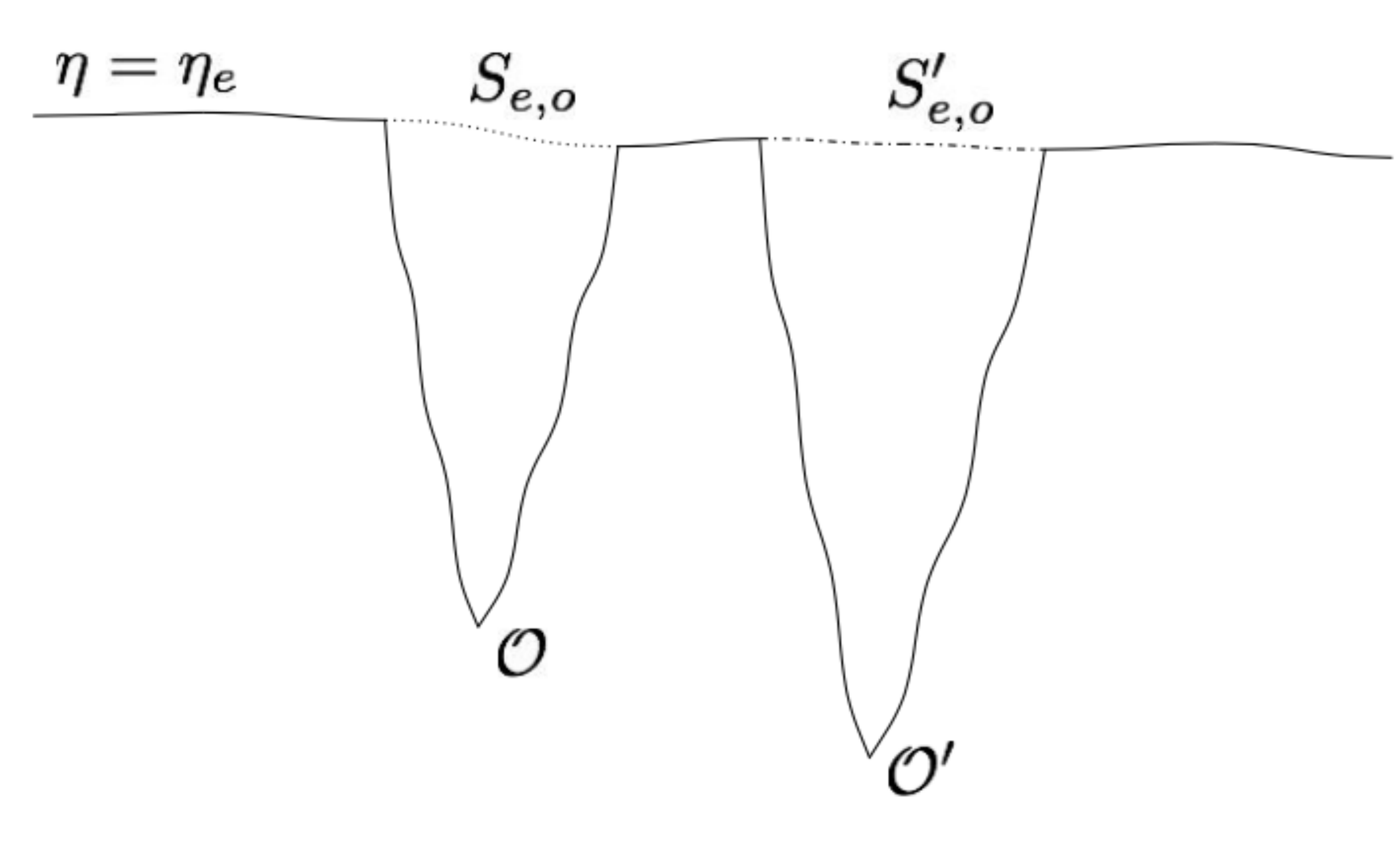}
	\caption{Different observers define different hypersurfaces $S_{e,o}$ each one with its own mean value temperature $ \bar{T}_{e}$. The deviation of $ \bar{T}_{e}$ from the mean temperature at the hypersurface $ \eta = \eta_e$ will therefore depend on the observer's position $ x_{o}$, and that information is encoded into $ \cT_o$.}
	\label{fig:lss}
\end{figure}
We stress that the mean $ \braket{}$ is taken on the space-like 3D-hypersurfaces of constant $ \eta$. 
However, what is important for the CMB is the mean taken on the last scattering surface%
\footnote{Again, $ \braket{T}_{e}$ is relevant for defining the time of emission $ \eta_e$.
} %
$S_{e,o}$. Here we define $S_{e,o}$ as the 2D-surface (usually thought as a deformed spherical shell) formed at the intersection between the hypersurface of $ \eta = \eta_e$ and the past light-cone of the observer. It follows from \eq{eq:ToTeExpI0-intro} that \textit{the observed mean temperature} is 
\begin{align}
	\bar{T}_o( x_{o};\eta_e) & = \f{a_e}{a_o} \braket{T}_{e} e^{ \cT_o} \,, 
	\label{eq:barTo}
\end{align}
where we have defined the intrinsic temperature anisotropies at the observer's spacetime position $ x_{o}$ as a mean value on the last scattering surface%
\footnote{Mean values on $S_{e,o}$ represent integrations w.r.t the direction of observation $ n_{o}$.
} %
$S_{e,o}$
\begin{align}
	e^{\cT_o} \equiv  \overline{ \exp \lp { \cT_{e} + \Phi_e - \Phi_o + I_{0} + 
	\ln \gamma_o \lp 1 - n_o \cdot v_o \rp - \ln \gamma_e \lp 1 - n_e \cdot v_e \rp} \rp} \,.
\end{align}
From \eq{eq:barTo} it follows that $ \cT_o = \cT_o ( x_{o};\eta_e)$ transforms under gauge transformations in the same way as the logarithmic anisotropies $ \cT_{e}$ but evaluated at the observer's position. Since this definition is valid for any observer with $ \eta_o  > \eta_e$, it provides a natural extension for the field $ \cT$ to the whole spacetime. Note however that by construction $ \cT_o$ depends only on the spacetime position $ x_{o}$ not on the direction of observation $ n_{o}$. This is in contrast with  intrinsic temperature anisotropies $ \cT_e$ which according to the discussion at the beginning of this section, could depend%
\footnote{In it does, as the intrinsic temperature anisotropies have a quadrupole component which act as a source for the CMB polarization \cite{Rees:1968,Ferte:2015oga}.
} %
on $ n_{e}$. It follows from Eqs.~\eqref{eq:ToTeExpI0-intro} and \eqref{eq:barTo} that \textit{the observed temperature} can be written as $ T_o = \bar{T}_o\ e^{\Theta}$, with
\begin{align}
	\Theta \equiv (\cT_e - \cT_o) + (\Phi_e - \Phi_o)  + I_0 + 
	%
	\ln \lp \f{\gamma_e \lp 1 - n_e \cdot v_e \rp}{\gamma_o \lp 1 - n_o \cdot v_o \rp} \rp \,.
	\label{eq:Gsw}
\end{align}
Finally, we define the mean temperature $ \bar{T}_e$ of the last scattering surface as
\begin{align}
	\bar{T}_e( x_{o};\eta_e) \equiv \braket{T}_{e} e^{ \cT_o} \,.
\end{align}
Because in general $ \bar{T}_e \ne \braket{T}_{e}$, then through the previous equation, $ \cT_o$ tell us how anisotropic the last scattering surface is (see figure~\ref{fig:lss}). From \eq{eq:barTo} it follows that
\begin{align}
	\bar{T}_o = \bar{T}_e \f{a_e}{a_o}\,.
\end{align}
The previous relation is simply the statement that the mean temperature evolves only through the cosmological expansion. Additionally, the quantity $ \Theta$ is what we call the logarithmic CMB temperature anisotropies, \eq{eq:Gsw} is the non-linear generalization of the Sachs-Wolfe formula, and $ I_0$ (given in \eq{eq:I0}) is the non-linear generalization of the integrated Sachs-Wolfe effect. As we will show in \refs{sec:SW1srt}, the presence of the factor $ \cT_o$ will guarantee the gauge invariance of $ \Theta$ and $ \bar{T}_{o}$. \eq{eq:Gsw} is the main result of this paper.

In the remaining sections we do the explicit calculations and consider particular cases. So for instance, in \refs{sec:review} give a quick review of fundamental concepts and introduce the notation. In \refs{sec:metric-tetrad} we introduce a tetrad basis which facilitates the resolution of the geodesic equation and allow us to interpret the metric components $ \beta_i$ as the tetrad components of the four-velocity of comoving observers. Then we compare our results with the previous ones in literature. Firstly, we consider the first-order case in \refs{sec:SW1srt} and discuss the gauge invariance. In \refs{sec:lensing} we show how to obtain the lensing term up to the desired order, and then in \refs{sec:SW2nd-curved} we obtain the second-order Sachs-Wolfe formula which is simpler than the previous ones in literature and then we give the conclusions. 
In a companion paper \cite{Roldan:2017kgd} we discuss the subtle issue of second-order gauge transformations on the CMB, prove the gauge invariance of our second-order formula and introduce the concept of a cosmological river-frame. Further applications of our results and comparison with existing ones will appear elsewhere \cite{Roldan:2017-2}.

\section{Quick review of fundamental concepts}
\label{sec:review}

In this section we quickly review some concepts which will be important to find the exact solution for the Sachs-Wolfe formula and at the same time give us a clear geometrical meaning of each terms in that formula.

\subsection{Tetrads}

An orthonormal \textit{dual tetrad} $ \eb{a}(x)$, is a set of \textit{dual vectors} $ \eb{a} \equiv \lb \eb{0}, \eb{1},\eb{2},\eb{3} \rb $ attached to each point $ x^{\mu}$ of the spacetime in which the line element looks Minkowskian%
\footnote{For an introduction to the tetrads we refer the reader to \cite{Ellis:1998ct,Carroll:1997ar}. Here, I am using the very nice notation used \cite{Pettinari:2014vja}.
} %
\begin{align}
	ds^{2} = \eta_{\underline{ab}}\ \eb{a}\ \eb{b}\,,
\end{align}
and so the tetrad axes form (at each point) a locally inertial orthonormal frame. We can transform between the tetrad frame and the coordinate frame by using the matrix $ \e{a}{b} $ and its inverse $ \e{a}{b}$,
\begin{align}
	\eb{a} = \e{a}{b}\ \dd x^{b} \,, \qquad  \dd x^{a} = \eI{b}{a}\ \eb{b} \,.
	\label{eq:tetradDef}
\end{align}

Now, the orthonormal \textit{tetrads} $ \ebI{a}$ (that is, the duals of $ \eb{a}$) are related to the coordinate vectors $ \p{a}$ (the duals of $ \dd x^{a}$) by
\OLD{The dual vectors of $ \dd x^{a}$ ($ \p{a}$) are also related to the dual vectors of $ \eb{a}$ ($ \ebI{a}$) by}
\begin{align}
	\ebI{a} = \eI{a}{b}\ \p{b} \,, \qquad  \p{a} = \e{b}{a}\ \ebI{b} \,.
\end{align}
Since any vector (or tensor) can be expressed in any base, we can write for instance (for a vector $ v$ and co-vector $ k$)
\begin{align}
	v = v^{a} \p{a} = v^{ \underline{b}}\ \ebI{b} \,, \qquad 
	k = k_{a} \dd x^{a} = k_{ \underline{b}}\ \eb{b}\,,
\end{align}
and by using the change of basis matrices, we can obtain the transformation rules for the components 
\begin{align}
	v^{a} = \un{v}{b}\ \eI{b}{a} \,, \qquad  
	k_{a} &  = \un{k}{b}\ \e{b}{a} \,, \notag \\
	\un{v}{b} = v^{ a}\ \e{b}{a} \,, \qquad  
	\unI{k}{b} &  = k_{a}\ \eI{b}{a} \,.
\end{align}
The same analysis can be made for tensors. In particular, for the metric tensor we have that the components transform as: $ g_{ \underline{ab}} = \eI{a}{\mu} \eI{b}{\nu} g_{\mu \nu}$, but we defined the tetrads to be orthonormal, in the sense that the metric looks Minkowskian (\eq{eq:tetradDef}), therefore $ g_{ \underline{ab}} = \eta_{ \underline{ab}}$, and we get
\begin{align}
	\eta_{ \underline{ab}} = \eI{a}{\mu} \eI{b}{\nu} g_{\mu \nu} \,, 
	\qquad g_{\mu \nu} = \e{a}{\mu} \e{b}{\nu} \eta_{ \underline{ab}}\,,
\end{align}
with similar expressions for the inverse matrices $ g^{ab}$ and $ \eta^{ab}$. Finally, since the metric $ g$ is used to rise and lower spacetime indexes, we can easily see that the metric $ \eta$ is used to rise and lower tetrad indexes, that is: $ \un{v}{a} = \eta^{ \underline{ab}}\ \unI{v}{b}\ $ and $ \unI{v}{a}\ = \eta_{ \underline{ab}}\ \un{v}{b}$.

\subsection{Photons and observers}
\label{sec:Photons-and-obs}

In this paper we use the signature $ -1$ for the metric. So, the four-velocity of a given observer satisfies $u \cdot u = - 1$, where a ``$ \cdot$'' represents the scalar product between four-vectors, that is, $ u \cdot u  = u^{a} u_{a} = \un{u}{a}\ \unI{u}{a}$. 

Two observers $ u_{2}$ and $ u_{1}$ are related by%
\footnote{Eqs.~\eqref{eq:v}-\eqref{eq:n} are given in a series of articles that follow the so-called 1 + 3 covariant approach to general relativity. See for instance \cite{King:1972td,Maartens:1998xg,Tsagas:2007yx}.
} %
\begin{align}
	u_{2}  & = \gamma_{(1),2} \lp u_1 + v_{(1),2} \rp \,, \qquad  \textrm{where} \notag \\
	u_1  & \cdot v_{(1),2} = 0 \,, \qquad 
	\gamma_{(1),2} = - u_{1} \cdot u_2 = \f{1}{ \sqrt{1 - v_{(1),2} \cdot v_{(1),2}}}\,, 
	\label{eq:v}
\end{align}
and $ v_{(1),2}$ is the relative velocity of $ u_{2}$ w.r.t $ u_1$. For an observer $ u^{a}$, the four-momentum $ p^{a}$ of given a photon can be written as%
\footnote{Note that we are writing scalars in capital letters and vectors and tensors in small letters.
} %
\begin{align}
	p = E \lp u - n \rp \,, \qquad  \textrm{with}
	\qquad u \cdot n = 0 \,, \qquad E = - p \cdot u\,, 
	\label{eq:n}
\end{align}
where $ E$ and  $ n^{a} $ are the observed energy and direction of arrival. Note that $ n \cdot n = 1$ and that $ d^{a} \equiv - n^{a}$ is the direction of propagation of the photon. In the following, it will be useful to introduce the concepts of comoving-observers $ u_{\textrm{com}}$ and tetrad-comoving observers $ \tilde{u}$, they are defined by the relations
\begin{align}
	\com{u}^{i}  & = 0 \,, \qquad \textrm{comoving-observers}\,, \\
	\un{ \tilde{u}}{i} & = 0 \,, \qquad \textrm{tetrad-comoving-observers}\,.
\end{align}
Note that in general, a comoving observer do not coincide with a tetrad-comoving observer. In fact, for the former the tetrads components of the four-velocity are $ \un{ \com{u}}{a} = \e{a}{0}\ u^{0}$, showing that in general $ \un{u}{i}$ do not vanish. For tetrad-comoving-observers the energy and direction of incoming photons has a simple form
\begin{align}
	\tilde{E}  & = - \unI{p}{0} \,, \qquad  \unI{ \tilde{n}}{a}  = \lp 0, \unI{p}{i}/ \unI{p}{0} \rp \,, 
	\label{eq:EtoCF}
\end{align}
additionally the decomposition of the four-velocity $ u = \tilde{\gamma} \lp \tilde{u} + \tilde{v} \rp $ is quite simple
\begin{align}
	\tilde{\gamma} = \un{u}{0} = \sqrt{1 + \un{u}{i}\ \unI{u}{i}} \,, 
	\qquad \un{ \tilde{v}}{a}  = \lp 0, \un{u}{i}/ \un{u}{0} \rp\,.
	\label{eq:vtoCF}
\end{align}
Using these results we can relate the energy $ E$ and direction $ n$ as observed by $ u$, with the energy $ \tilde{E}$ and direction $ \tilde{n}$ as seen by tetrad-comoving-observers simply by
\begin{align}
	E & = \tilde{E}\lp \un{u}{0} + \tilde{n} \cdot u \rp \,,
	\label{eq:EinTCF}\\
	 \un{\tilde{n}}{i} & = \f{ \un{n}{i} - \un{u}{i}}{\un{u}{0} - \un{n}{0}}\,.
	 \label{eq:ninTCF}
\end{align}
For comparison with other works in the literature, let's now relate the observed energy $ E$ to the energy seen by comoving observers $ \com{E}$. We can obtain two equivalent expressions: the first one is obtained by applying \eq{eq:EinTCF} two times%
\footnote{That is, we apply two boost: one from the comoving frame to the tetrad-comoving one, and then one additional boost to the u-observer frame. 
} %
\begin{align}
	E = \com{E} 
	\f{\lp \un{u}{0} + \tilde{n} \cdot u \rp}{\lp \un{\com{u} }{0}\  + \tilde{n} \cdot \com{u} \rp}\,, 
\end{align}
and the other one follows by applying the boost directly from the comoving observer to the u-observer 
\begin{align}
	 E = \f{\com{E}}{ \gamma \lp 1 + n \cdot \com{v} \rp } = 
	 \f{\com{E}}{ \gamma \lp 1 - n \cdot v \rp }\,, 
	\label{eq:EtoEcom}
\end{align}
where $ \com{v}$ is the velocity of $ \com{u}$ w.r.t $u$, and we have introduced the \textit{peculiar velocity} $ v \equiv  - \com{v}$, that is, $ \com{u} = \gamma \lp u - v \rp $. Although we will call $ v$ the peculiar velocity of the observer, it is clear that this is not the velocity of $ u$ with respect to $ \com{u}$. We have introduced this concept in order to be closer to the notation used in many other works, see for instance Eq.(1) of \cite{Aghanim:2013suk}. It follows from the previous equations that
\begin{align}
	\f{1}{ \gamma \lp 1 - n \cdot v \rp } = 
	\f{\lp \un{u}{0} + \tilde{n} \cdot u \rp}{\lp \un{\com{u} }{0}\ + \tilde{n} \cdot \com{u} \rp}\,.
	\label{eq:nTOncom}
\end{align}
Although the photons's energy has a simply form in the comoving frame,
\begin{align}
	 \com{E} = - p_{0} \com{u}^{0} = - \f{p_{0}}{ \sqrt{g_{00}}}\,,
\end{align}
most of the time we prefer to work with tetrad-comoving-observers because of the nice properties given in this frame (see Eqs.~\eqref{eq:EtoCF}-\eqref{eq:vtoCF}). In particular, for the direction $ \tilde{n}$ we have $ \un{n}{a} = \unI{n}{a}$ making it safe to use bold-notation (see below). 
By contrast, in the comoving frame we have $ (\com{n})_{0} = 0$ but in general $ \com{n}^{0} \ne 0$. Additionally, the physics becomes more transparent when using locally orthonormal basis (tetrads) instead of coordinates  basis.

\subsection{Observed CMB temperature}

It is well known (see for instance pp.588 of \cite{Misner:1974qy}) that in absence of secondary scatterings the CMB temperature at the point of observation $ T_{o}$ is related with the temperature at emission $ T_{e}$ by (this is a consequence of the Liouville theorem)
\begin{align}
	T_{o} = \frac{E_o}{E_e} T_e \,,
	\label{eq:ToTe}
\end{align}
where $ E_{e}$ ($ E_{o}$) is the energy of photons at the emission (observation) point. Note that in general, the temperature is a function of both: the spacetime position $ x$ and direction $ n$
\begin{align}
	T_{o} = T( x_o,n_o)  \,, \qquad  T_{e} = T( x_e, n_e) \,.
\end{align}
The direction of emission (as seen by a local observer) is $ d_e =  - n_e$. Note that before the period of recombination it is expected that the temperature of the photon fluid is isotropic, in that sense it will not depend on the direction of emission. However, during the period of recombination a small quadrupole anisotropy arise in the photon distribution function \cite{Kaiser:1983,Hu:1995em}, that is why we kept the $ n_e$ dependence in the emission temperature.

By using Eqs.~\eqref{eq:EtoCF} and \eqref{eq:EinTCF}, the observed temperature can be written as
\begin{align}
	T_{o} = T_{e} \f{ \unI{p}{0}( x_o) }{\unI{p}{0}(x_e)} 
	\f{ \lp \un{u}{0} + \tilde{n} \cdot u \rp_o  }{\lp \un{u}{0} + \tilde{n} \cdot u \rp_e}\,,
	\label{eq:ToTe-tetrad}
\end{align}
This can also be written in bold notation as
\begin{align}
	T_{o} = T_{e} \f{ \unI{p}{0}( x_o) }{\unI{p}{0}(x_e)} 
	\f{ \sqrt{1 + \vu_o^{2}} + \vnt_o \cdot \vu_o }{\sqrt{1 + \vu_e^{2}} + \vnt_e \cdot \vu_e}\,,
	\label{eq:ToTeBold}
\end{align}
where the bold notation is used as a shorthand to express the spatial components in the tetrad basis, that is $ \vnt = ( \un{ \tilde{n}}{i})$, $ \vu = ( \un{u}{i})$, and $ \vu \cdot \vnt =  \unI{u}{i}\ \un{ \tilde{n}}{i}$. We stress that $ \vnt$ is the observed direction of incoming photons as seen by the tetrad-comoving-observers, which is related to the direction of observation $ \vn$ by \eq{eq:ninTCF}. The previous equation is equivalent to that given in appendix A of \cite{Mirbabayi:2014hda}, although there the authors were only interested in obtaining the CMB temperature up to second order in the Poisson gauge, and by neglecting primordial vector and tensor perturbations. In this paper however, we will not use the bold-notation.

\section{The metric and tetrad components}
\label{sec:metric-tetrad}

In this section we introduce the metric and tetrads which will allow us to obtain the Sachs-Wolfe formula. Note that two common notations for the metric are
\begin{align}
	\dd s^2  & = a^2(\eta) \lc - \lp 1 + 2\phi\rp \dd \eta^2 + 
	2\omega_{i}\ \dd x^i \dd \eta+ \lb (1 - 2 \psi) \delta_{ij} + 2\gamma_{ij}\rb  \dd x^i \dd x^j \rc \,, 
	\label{eq:metric1}\\
	 & = a^2(\eta) \lc - e^{2\Phi} \dd \eta^2+ 2\omega_{i}\ \dd x^i \dd \eta + 
	 \lb  e^{-2\Psi} \delta_{ij} + 2\gamma_{ij}\rb  \dd x^i \dd x^j \rc \,, 
	\label{eq:metric2}
\end{align}
where $ \delta_{ij}$ is the delta Kronecker tensor, $x^\mu=( \eta, x^{i})$, $\eta$ the conformal time, $a$ is the scale factor and $\gamma_{ij}$ is defined as traceless in order to make the separation of the spatial part of the metric unambiguous. Usually each quantity is expanded perturbatively into first, second, or third order perturbations. Here however, we will treat each quantity non-perturbatively. We propose to use the following parametrization of the metric
\begin{align}
	\dd s^2  & = a^2(\eta) \lc - e^{2\Phi} \dd \eta^2 + 
	2\beta_{j} \lp e^{ \Phi - \Psi - \Gamma} \rp_{ji} \ \dd x^i \dd \eta + 
	\lp  e^{-2\lp \Psi+ \Gamma\rp}\rp_{ij}\dd x^i \dd x^j \rc \notag \\
	 &  = a^2(\eta) e^{2\Phi} \dd \hat{s}^2 \,,
	\label{eq:Metric-CMetric}
\end{align}
where $ \Gamma$ is a symmetric and traceless matrix, and the notation $ \Psi+ \Gamma$ really means $ \Psi 1 + \Gamma$ where $ 1$ is the identity matrix, that is, $ \lp \Psi+ \Gamma \rp_{ij} = \Psi \delta_{ij} + \Gamma_{ij}  $. The conformal metric is
\begin{align}
	\dd \hat{s}^2 \equiv - \dd \eta^2 + 
	2\beta_{j} \lp e^{-M} \rp_{ji} \ \dd x^i \dd \eta + \lp  e^{-2M}\rp_{ij}\dd x^i \dd x^j \,,
	\label{eq:Cmetric}
\end{align}
with $ M = \Phi + \Psi + \Gamma$. Note that indexes in $ \beta_i$ and $ M_{ij}$ are raised and lowered with $ \delta_{ij}$.

Hereafter, we will mainly work with the conformal metric \eq{eq:Cmetric}, and whenever we need to express quantities in the physical metric we just multiply by the appropriated conformal factor, as given for instance in \eq{eq:Metric-CMetric} (more details below).

We will still rewrite the conformal metric in a different way that will allow us to give an interesting interpretation of $ \beta_i$ and to easily express the metric in terms of tetrads,%
\footnote{This is basically the ADM decomposition of the metric.
} %
\begin{align}
	\dd \hat{s}^2 = - \lp \beta^0\dd \eta \rp ^2 + 
	\lc \lp e^{ - M} \rp ^{j}_{\ i} \dd x^i + \beta^j \dd \eta\rc 
	\lc \lp e^{ - M} \rp_{ jk} \dd x^k + \beta_j \dd \eta\rc \,,
	\label{eq:GPmetric}
\end{align}
where we have introduced $ \beta^0 \equiv \sqrt{1 + \beta_i \beta^i} $. It is interesting to note that null paths in the conformal s-t are%
\footnote{Hereafter we will use ``s-t'' as a short-hand for spacetime.
} %
also null paths in the physical s-t, that implies that the path of photons is totally determined by just two quantities: $ \beta_i$ and $ M_{ij}$ (and its derivatives, which enter the geodesic equation). This is  important for effects like lensing, time-delay and the integrated Sachs-Wolfe (ISW). %
%
%
%
%
The conformal metric in the form given in \eq{eq:GPmetric} provides a natural basis of orthonormal dual vectors $ \eb{a} = \e{a}{\mu}\ \dd x^{ \mu}$, whose tetrads components are
\begin{align}
	\e{0}{ \mu} = \beta^0 \delta^{0}_{\mu} \,,   \qquad  
	\e{i}{0} = \beta^i \,,  \qquad \e{i}{j} = \lp e^{-M} \rp ^{i}_{\ j}\,. 
	\label{eq:tetrads}
\end{align}
The tetrads for the physical s-t are obtained from the above ones, simply multiplying by the conformal factor $ a e^{ \Phi}$. We now note that $ \beta^{a}$ are the tetrad components of the four-velocity of comoving-observers. In fact, for a comoving observer ($ \com{u}^{i} = 0$) we have
\begin{align}
	\un{\com{u}}{a}\  = ( a e^{\Phi})\e{a}{0}\ \com{u}^{0} = 
	\e{a}{0} = \beta^{a}\,,
\end{align}
where we have multiplied by the conformal factor $ a e^{\Phi}$ in order to get quantities in the physical s-t. Additionally, we used the normalization condition to obtain $ a e^{\Phi} \com{u}^{0} = 1$.

Below, we provide some relations which will be useful in the next section. They are the inverse tetrads, 
\begin{align}
	\eI{a}{0} = \f{1}{\beta^0} \delta^{0}_{a} \,, \qquad 	
	\eI{0}{i} = - \f{1}{\beta^0} \lp e^{M} \rp^{i}_{\ j}\ \beta^{j} \,,   \qquad  
	\eI{j}{i} = \lp e^{M} \rp ^{i}_{\ j}\,,
	\label{eq:Itetrads}
\end{align}
and the derivative of the exponential matrix, the Baker-Campbell-Hausdorff formula (or the Zassenhaus formula) \cite{Wilcox:1967zz}
\begin{align}
	\lp \p{\mu} e^{-M} \rp = - A_{\mu}\ e^{-M} \,, 
	\qquad A_{\mu} \equiv \int_{0}^{1} \dd s\ e^{- s M} \lp \p{\mu} M \rp e^{s M}\,.
	\label{eq:Adef}
\end{align}
\OLD{SEE \url{https://www.wikiwand.com/en/Matrix_exponential#/cite_ref-3}, }

\subsection{The geodesic equation}
\label{sec:geo-eq}

In order to obtain the explicit form of $ \tilde{n}$ and $ \unI{p}{0}$ needed to obtain the observed CMB temperature in \eq{eq:ToTe-tetrad}, we need to solve the geodesic equation. Since photons follow null paths, we can use the conformal s-t instead of the physical s-t, this will make calculations easier. Note that, if $ p^\mu$ is the photon four-momentum in physical s-t, then $ \hat{p}^{\mu} = (a e^{\Phi})^{2} p^{\mu}$ is the photon four-momentum in conformal s-t (Appendix D of \cite{Wald:1984rg}), consequently, $ \unI{ \hat{p}}{a} = (a e^{\Phi}) \unI{p}{a}$. Note also that according to \eq{eq:EtoCF}, the direction of observation as seen in the tetrad-comoving-frame is $ \unI{\tilde{n}}{i} =  - \unI{p}{i}/\unI{p}{0} =  - \unI{\hat{p}}{i}/\unI{\hat{p}}{0}$. With those considerations in mind, we can now proceed to obtain the observed CMB temperature. We start with the geodesic equation in the conformal s-t \cite{Carroll:1997ar}
%
%
\begin{align}
	\frac{d \hat{p}_\mu}{d\lambda} = \f12 \lp \p{\mu} \hat{g}_{\alpha\beta} \rp \ 
	\hat{p}^\alpha \hat{p}^\beta  = \lp \p{ \mu} \e{a}{ \nu}\rp \eu{b}{\nu}\ 
	\unI{\hat{p}}{a}\ \unI{\hat{p}}{b}\,, 
\end{align}
where $ \lambda$ is an affine parameter. Using $ \f{d}{d \lambda} = \hat{p}^{0} \f{d}{d \eta}$, and after dividing on each size by $ \lp \unI{\hat{p}}{0} \rp^{2}$ we get
\begin{align}
	 - \f{1}{\beta^{0}}\frac{ \dot{ \hat{p}}_\mu}{ \unI{\hat{p}}{0}} =  
	 \lp \p{ \mu} \e{a}{ \nu}\rp \eu{b}{\nu}\ \f{\unI{\hat{p}}{a}}{\unI{\hat{p}}{0}}\ 
	 \f{\unI{\hat{p}}{b}}{\unI{\hat{p}}{0}}\,, 
\end{align}
where we used $ \unI{\hat{p}}{0} = - \beta^{0} \hat{p}^{0}$, and a ``dot'' over a variable means total derivative w.r.t conformal time. By noting that $ \hat{p}_{0} = \beta^{0} \unI{\hat{p}}{0} + \beta^{i} \unI{\hat{p}}{i} $, we can write
\begin{align}
	\f{1}{\unI{\hat{p}}{0}} = \f{1}{\hat{p}_{0}} \lp \beta^{0} + \beta^{i} \f{\unI{\hat{p}}{i}}{\unI{\hat{p}}{0}} \rp \,,
	\label{eq:po-p0}
\end{align}
and therefore the geodesic equation takes the form
\begin{align}
	\frac{ \dot{ \hat{p}}_\mu}{ \hat{p}_{0}} = \f{1}{\beta^{0} + \beta^{i} \unI{\tilde{n}}{i}}
	\lb \lp \p{\mu} \beta^{0} \rp + \un{\tilde{n}}{i} \lc 
	\lp \p{\mu} \beta_{i} \rp + \lp A_{ \mu} \rp_{ij}\ \beta^{j} \rc  + 
	\beta^{0}\un{\tilde{n}}{i} \lp A_{ \mu} \rp_{ij} \un{\tilde{n}}{j} \rb\,, 
	\label{eq:geoEq}
\end{align}
where we have used \eq{eq:Adef} for $ A_{ \mu}$. The equation above can be integrated for $ \mu = 0$, yielding 
\begin{equation}
\begin{split}
	 \hat{p}_{0} (x_o)  & = \hat{p}_{0} (x_e) e^{I_{0}}\,, \\
	 I_{0} & \equiv  \int_{ \eta_e}^{ \eta_o} \f{\dd\eta}{\beta^{0} + \beta^{i} \unI{\tilde{n}}{i}}
	\lb \lp \p{0} \beta^{0} \rp + \un{\tilde{n}}{i} \lc \lp \p{0} \beta_{i} \rp + 
	\lp A_{ 0} \rp_{ij}\ \beta^{j} \rc  + \beta^{0}\un{\tilde{n}}{i} 
	\lp A_{ 0} \rp_{ij} \un{\tilde{n}}{j} \rb\,, 
	\label{eq:I0-1}
\end{split}
\end{equation}
which after substituting into \eq{eq:ToTe-tetrad} yields (multiplying by the conformal factor)
\begin{align}
	T_{o} = T_{e} \f{ \lp a e^{\Phi} \rp_e }{\lp a e^{\Phi} \rp_o}\ e^{I_{0}}\ \f{ \lp \beta^{0} + 
	\un{ \tilde{n}}{i}\ \beta_{i} \rp_e }{\lp \beta^{0} + \un{ \tilde{n}}{i}\ \beta_{i} \rp_o}\ 
	\f{ \lp \un{u}{0} + \tilde{n} \cdot u \rp_o  }{\lp \un{u}{0} + \tilde{n} \cdot u \rp_e}\,,
	\label{eq:ToTeExpI0-ini-Eq}
\end{align}
where we have used \eq{eq:po-p0}. Note that \eq{eq:po-p0} is nothing else that the relation between the energy in the comoving frame $ \com{E} = - p_0$, and the energy in the tetrad-comoving-frame $ \tilde{E} = - \unI{p}{0}$, that is, 
\begin{align}
	\com{E} & = \tilde{E} \lp \un{\com{u} }{0}\  + \tilde{n} \cdot \com{u} \rp \,,
\end{align}
which follows from \eq{eq:EinTCF}. We see that $ \beta^{0} + \un{ \tilde{n}}{i}\ \beta_{i} = \un{\com{u} }{0}\  + \tilde{n} \cdot \com{u}$ represents a Doppler boost. This is however, a point-to-point (along the photon's path) boost which takes the observed temperature by tetrad-comoving-observers into the observed temperature by comoving observers. On the other hand, since this boost is determined by $ \beta^{i}$, which is directly related to the $ 0-i$ components of the metric, we will call this ``a metric-Doppler effect''.

By using \eq{eq:ToTeExpI0-ini-Eq} together with \eq{eq:nTOncom} we can equivalently write 
\begin{align}
	T_{o} = T_{e} \f{a_e}{a_o}\ e^{\Phi_e - \Phi_o + I_{0}}\ 
	\f{\gamma_e \lp 1 - n \cdot v \rp_e}{\gamma_o \lp 1 - n \cdot v \rp_o}\,.
	\label{eq:ToTeExpI0}
\end{align}
This is the equation we used in \refs{sec:gSW} to obtain the generalized Sachs-Wolfe formula \eq{eq:Gsw}.

To complete the solution for the observed temperature, we need to obtain both $ \un{\tilde{n}}{i}$ and the coordinates $ x^{\mu}$ of the photon's path. We will address this problem in \refs{sec:lensing}. Finally, we can write the ISW in a covariant way by noting that 
\begin{align}
	\un{ \tb}{a}  & \equiv \lp 0, \beta^{i}/\beta^{0} \rp \,,
	\label{eq:tb}
\end{align}
is the velocity of comoving observers w.r.t tetrad-comoving-observers (it follows from \eq{eq:vtoCF}). In terms of $ \tb$ we have
\begin{align}
	I_{0} & \equiv  \int_{ \eta_e}^{ \eta_o} \dd \eta\ \lc \f{ \tb \cdot \tb'}{1- \tb^{2}} + 
	\f{ \tn \cdot \tb' + \tn \cdot A_{0} \cdot \lp \tn + \tb \rp }{1 + \tb \cdot \tn}\rc \,.
	\label{eq:I0}
\end{align}
Here, we are treat $ (A_{0})_{ij}$ as the non-vanishing components of a (space-like) tensor $ A_{0}$ in the tetrad-frame, that is: $ \unI{(A_{0})}{ij} \equiv (A_{0})_{ij}$ and $ \unI{(A_{0})}{0 a} = 0$.

\OLD{The advantage of using tetrad-comoving-observers instead of comoving-observers is that for the former we have $ \un{\tilde{n}}{0} = 0$ so it is meaningful to use the bold-notation $ (\un{\tilde{n}}{i}) = \vnt$. For the latter, the bold-notation is not appropriated as $ \beta^{0} \ne 0$. 
}

\subsection{Sachs-Wolfe at first order}
\label{sec:SW1srt}

In this section we use the generalized Sachs-Wolfe formula \eq{eq:Gsw} to obtain the well known results at first order. Then we will discuss the gauge invariance of our result, emphasizing the importance of the factor $ \cT_o$.

From the definition of $ A_{\mu}$, \eq{eq:Adef}, we have up to first order $ A_{\mu} = \p{\mu} M $. Additionally, since we are interested in writing the observed temperature up to first order, we can take $ \tilde{n}$ at zero-order as it is always multiplying first-order quantities. That has several consequences. i) We can drop the ``tilde'' in the direction of observation as it is the same (at zero-order) for all observers, that is, $ \tilde{n} = \com{n}$ = n, ii) all quantities are evaluated along the unperturbed path for which
we can set $ \un{n}{i}_o = \un{n}{i}_e  = \un{n}{i}$. This is called the Born approximation, and the unperturbed path has coordinates $ x^{i}( \eta) = x^{i}_{o} + \lp \eta_o-\eta\rp n_o^{i}$. Under these considerations the Sachs-Wolfe formula up to first order is
\begin{align}
	\Theta  & = (\cT_e - \cT_o) + (\Phi_e - \Phi_o)  + I_0 +  
	\ln \lp \f{ 1 - n_e \cdot v_e }{1 - n_o \cdot v_o} \rp \,, 
	\label{eq:To1st-0}\\
	 I_{0} & = \int_{ \eta_e}^{ \eta_o} 
	\lb \un{n}{i}\ \beta_i' + \un{n}{i} M_{ij} \un{n}{j} \rb\,, 
\end{align}
where a ``prime'' means partial derivative w.r.t conformal time, and we have used $ \beta^0  = 1 = \gamma$ valid up to first order.

Although the components of $ n_{o}$ are equal to the components of $ n_{e}$, that is, $ \un{n}{i}_o = \un{n}{i}_e$, we have written $ n_e \cdot v_e$ instead of $ n_o \cdot v_e$ because in general, the quantity $ n_o \cdot v_e$ is not well defined as it represents the scalar product of two four-vectors which are defined at different points in the s-t.

Remembering that $ M_{ij} = (\Phi + \Psi) \delta_{ij} + \Gamma_{ij}$ (see after \eq{eq:Cmetric}), and expanding the logarithm up to first order we obtain
\begin{align}
	\Theta  & = (\cT_e - \cT_o) + (\Phi_e - \Phi_o) - ( v_e \cdot n_e  - v_o \cdot n_o) + I_{0}\,, 
	\label{eq:SW1rst}	\\
	 I_{0} & = \int_{ \eta_e}^{ \eta_o} 
	\lb \un{n}{i}\ \beta_i' + \Phi' + \Psi' + \un{n}{i} \Gamma_{ij}' \un{n}{j} \rb\,.
\end{align}
This is (apart from the factor $ \cT_o$) the very well known first-order Sachs-Wolfe formula given in \eq{eq:SW}.

\subsection{Gauge invariance}


We now discuss the gauge invariance of \eq{eq:SW1rst}. The gauge invariance of our results up to second order are discussed in a companion paper \cite{Roldan:2017kgd}. There we will provide the full set of transformation rules for the metric components and additional relevant quantities. We will use the following notation: under a gauge transformation a geometrical object  $ T$ (scalar, vector, tensor, connections, etc.) will transform as $ T \to T  + \Delta T$. Here we just need the first-order gauge transformations induced by the gauge generator $\xi^\mu=( \alpha, \xi^{i})$, so we have%
\footnote{These transformation rules can also be obtained easily from the rules given in \cite{Malik:2008im,Matarrese:1997ay}.
} %
\begin{align}
	\Delta \cT & = - \cH \alpha \,, \qquad \quad
	\Delta \Phi = \alpha'+\cH\,\alpha \,, \qquad \quad
	\Delta v_{i} = - \xi_{i}'\,,\\
	\Delta \beta_i & =\xi_i'- \alpha_{,i} \,, \qquad \qquad
	\Delta M_{ij} = \alpha' \delta_{ij}  - \xi_{(i,j)}\,, 
	%
\end{align}
where $ \cH \equiv a'/a$ is the Hubble's expansion rate, a ``comma'' means derivative, so that $ \alpha_{,i} = \p{i} \alpha$. The parenthesis in the expression $ \xi_{(i,j)}$ means symmetrization, so $ \xi_{(i,j)} = (\xi_{i,j}+\xi_{j,i})/2$.

With these expressions, it is easy to show the gauge invariance of $ \Theta$, that is $ \Delta\Theta = 0$. Indeed, for the integrated Sachs-Wolfe term, we get
\begin{align}
	\Delta I_{0} & = \int_{ \eta_e}^{ \eta_o} 
	\lb \lp \alpha'' -  \un{n}{i}\ \p{i} \alpha'\rp  + \un{n}{j} \lp \xi_{j}'' - 
	\un{n}{i}\ \p{i} \xi_{j}' \rp \rb \notag\\
	 & = \lp \alpha' + \un{n}{j}\ \xi'_{j} \rp \Big|_{e}^{o} \,, 
\end{align}
where we made used of the fact that along the unperturbed path, the following relations holds $ \p{0} - \un{n}{i}\ \p{i} = \dd/\dd \eta$. Additionally, we have
\begin{align}
	\Delta \lb  (\cT_e - \cT_o) + (\Phi_e - \Phi_o) - ( v_e \cdot n_e  - v_o \cdot n_o)\rp = 
	\lp \alpha' + \un{n}{j}\ \xi'_{j} \rb \Big|_{o}^{e}\,,
\end{align}
showing explicitly that $ \Theta$ is gauge invariant. Since the full temperature $ T_{o} = \bar{T}_{o} e^{ \Theta}$ is an observable, it has also to be gauge invariant, as a consequence the mean value $ \bar{T}_{o}$ also is. Note that this result was possible thanks to the presence of $ \cT_o$ inside $ \Theta$. Without it, each time we perform a gauge transformation, the temperature anisotropies would acquire an additional monopole term.

\section{The lensing term}
\label{sec:lensing}

To complete our analysis we need to obtain $ \un{\tilde{n}}{i}$ and the coordinates $ x^{\mu}$ along the photon's path. These quantities are needed for a fully computation of the ISW effect. Additionally, they provide the so called lensing and time-delay terms (see \refs{sec:SW2nd-Born}). In this section we arrive at an expression which can by solved easily by iteration, allowing us to obtain the solution perturbatively up to the desired order.

By manipulating \eq{eq:geoEq}, we can obtain a differential equation for $ \un{n}{i}$. We however choose to follow a different way which yields a compact expression and can be used to easily obtain the coordinates of the photon's path. We start by defining $ q^{a}$ (which is not a four-vector) by the relation
\begin{align}
	q^{a} \equiv \f{\hat{p}^{a}}{\hat{p}^{0}} = \f{ p^{a}}{p^{0}}  \,, \quad \Rightarrow \quad 
	q^{a} = \der{x^{a}}{\eta} = \lp 1, \dot{x}^{i} \rp \,, 
	\label{eq:qDEF}
\end{align}
and we remind the reader that a ``hat'' means that quantities belong to the conformal s-t. Now, by using $ \un{\tilde{n}}{i}\ =  - \un{p}{i}/ \un{p}{0} = - \e{i}{a}\ p^{a} / (\beta^{0}\ p^{0})$
%
%
%
%
we get
\begin{align}
	\un{\tilde{n}}{i}\ =  - \f{1}{ \beta^{0}} \lp \beta^i + \lp e^{-M} \rp^{i}_{\ j}\ q^{j} \rp \,.
	\label{eq:ni-qi}
\end{align}
Then if we manage to obtain $ q^{i}$, we automatically get both $ \un{n}{i}$ and $ x^{i} = \int \dd \eta\ q^{i}$. Therefore, we now focus on $ q^{i}$. Before proceeding, we stress that the previous relation is nothing else that transformation of the direction vector, from the comoving-frame to the tetrad-comoving-observers, \eq{eq:ninTCF}. That is, the previous relation can be written as
\begin{align}
		 \un{\tilde{n}}{i} & = \f{ \un{ \com{n}}{i} - \un{ \com{u}}{i}}{\un{ \com{u}}{0} - \un{ \com{n}}{0}}\,.
\end{align}

Consider now the geodesic equation in conformal s-t
\begin{align}
	\hat{p}^{0} \dot{ \hat{p}}^{a} + \cris{a}{b}{c}\ q^{b} q^{c}\ = 0\,,
\end{align}
then by using $ \dot{ \hat{p}}^{a}/\hat{p}^{0} = \dot{q}^{a} + q^{a}\ \dot{ \hat{p}}^{0}/ \hat{p}^{0} $ we get
\begin{align}
	 - \dot{q}^{a}  & = q^{b}\ \cris{a}{b}{c}\ q^{c} - q^{a} \lp q^{b}\ \cris{0}{b}{c}\ q^{c} \rp 
	 \notag \\
	  & = 
	q \cdot \Gamma^{a} \cdot q - q^{a}\ \lp q \cdot \Gamma^{0} \cdot q \rp  \,,
	\label{eq:qa}
\end{align}
where for simplicity of notation we have written on the second line $ q^{b}\ \cris{a}{b}{c}\ q^{c} = q \cdot \lp \Gamma^{a} \rp  \cdot q $, that is, we treat $ \cris{a}{b}{c}$ as the components of a matrix $ \Gamma^{a}$. The relevant part of \eq{eq:qa} is that for the spatial indices $ a = i$ and the $ a = 0$ component is automatically satisfied, with $ q^{0} = 1$. \eq{eq:qa} is a \textit{autonomous} cubic equation in $ q$, without an obvious analytic solution.%
\footnote{I thank Yves Daoust user from stackexchange.com for useful comments on this point. See: \newline 
 \url{https://math.stackexchange.com/questions/2205149/non-linear-matrix-differential-equation}
} %
This can however easily be solved perturbatively, so for instance, if we call $ q^{a}_{(n)}$ the solution up to n-order, we can immediately obtained (n + 1)-solution as
\begin{align}
	 - q^{a}_{(n + 1)} \Big|_{ \eta}^{ \eta_o} = \int_{ \eta}^{ \eta_o} \dd \eta\
	 q_{(n)} \cdot \lp \Gamma^{a} - q^{a}_{(n)}\ \Gamma^{0}\rp  \cdot q_{(n)} \,.
	\label{eq:qa-1}
\end{align}
We now detail the first-order solution which is needed to obtain the second-order logarithmic temperature anisotropies.

\subsection{Lensing term at first order}

As described before, we can just use \eq{eq:qa-1} to easily obtain the first-order solution for $ q^{a}$. Before doing the integration, however, let's write down the integrand on the r.h.s of \eq{eq:qa} in a suitable way. Let's start with 
\begin{align}
	\cris{a}{b}{c}\ q^{b} q^{c} & = \f12 \hat{g} ^{a \mu} \lp  - \hat{g}_{bc,\mu} + 
	\hat{g}_{\mu b,c} + \hat{g}_{c\mu,b} \rp q^{b} q^{c} \notag \\
	 & = - \f12  \lp \hat{g} ^{a \mu} \hat{g}_{bc,\mu} \rp q^{b} q^{c} + 
	 \hat{g} ^{a \mu} q^{b} \dot{\hat{g}}_{b \mu}\,,
	\label{eq:qa-1st-Deq}
\end{align}
where we have used the fact that $ q^{c} \p{c} = \dd/ \dd \eta$. Now, since $ \hat{g}_{ab,c}$ is already first order we can set $ \hat{g}^{ab}= \eta^{ab}$ on the previous equation, so we got from \eq{eq:qa}
\begin{align}
	 - \dot{q}^{j} & =  - \f12 q^{b} q^{c} \lc \hat{g}_{bc,j} + \hat{g}_{bc,0}\ q^{j} \rc 
	 - q^{b} \lp \dot{\hat{g}}_{jb} + \dot{\hat{g}}_{0b}\ q^{j} \rp \,, 
\end{align}
then noting that $ \p{0} = \dd/ \dd \eta - q^{i} \p{i}$, and defining the transverse gradient as
\begin{align}
	\p{i}^{\perp}  = \p{i} - \unI{\tilde{n}}{i}\ \lp \tilde{n} \cdot \p{} \rp \,, 
	\qquad \textrm{where} \qquad  \tilde{n} \cdot \p{} \equiv \un{\tilde{n}}{i}\ \p{i}\,,
\end{align}
we arrive, after integration, to
\begin{align}
	- q^{i} \Big|_{ \eta}^{ \eta_o} = 
	\lb \un{ \tb}{i} + 2 \un{\lp M \cdot \tilde{n} \rp}{i} - 
	\un{\tilde{n}}{i}\ \lp \tilde{n} \cdot M \cdot \tilde{n} \rp +  
	\int \dd \eta\ \p{i}^{\perp} \lp \tilde{n} \cdot \tb + 
	\tilde{n} \cdot M \cdot \tilde{n} \rp \rb \Big|_{ \eta}^{ \eta_o} \,.
	\label{eq:qi-1st-sol}
\end{align}
Here we have used that up to second order $ \beta^{i} = \un{ \tb}{i}$ (see \eq{eq:tb}), and we treat
$ M_{ij}$ as the non-vanishing components of a (space-like) tensor $ M$ in the tetrad-frame, that is: $ \unI{M}{ij} \equiv M_{ij}$ and $ \unI{M}{0 a} = 0$. This is pretty much the same as we did in \eq{eq:I0} for $ A_{0}$.

It is clear that on the r.h.s of the previous equation we should keep $ \tilde{n}$ at zero order, this fact was taken into account in passing from \eq{eq:qa-1st-Deq} to \eq{eq:qi-1st-sol} by setting $ q^{i} = - \un{ \tilde{n}}{i}$ valid at zero-order. Additionally at zero-order we have $ \un{\tilde{n}}{i}_{e} = \un{\tilde{n}}{i}_{o} = \un{\tilde{n}}{i}$ and we can also remove the tilde from $ \tilde{n}$, so that $ \un{\tilde{n}}{i} = \un{n}{i}$.

Now that we are in possession of $ q^{a}$, we can immediately obtain $ \tilde{n}$ and $ x^{a}$ up to first order.

\paragraph{Direction vector $ \tilde{n}$ up to first order}

In order to obtain $ \tilde{n}$, we see from \eq{eq:ni-qi} that up to first order $ \un{\tilde{n}}{i} = - \lp \un{ \tb}{i} + q^{i} + M_{ij}\ \un{\tilde{n}}{j} \rp$, so we get
\OLD{\begin{align*}
	\un{\tilde{n}}{i} \Big|_{ \eta}^{ \eta_o} = 
	\lb \lp M \cdot \tilde{n} \rp^{i} - \un{\tilde{n}}{i}\ \lp \tilde{n} \cdot M \cdot \tilde{n} \rp %
	 + \int \dd \eta\ \p{i}^{\perp} \lp \tilde{n} \cdot \beta + \tilde{n} \cdot M \cdot \tilde{n} \rp \rb
	  \Big|_{ \eta}^{ \eta_o} 
	 \,,
\end{align*}
}
\begin{align}
	\un{\tilde{n}}{i} = \un{\tilde{n}}{i}_{o} - 
	\lc \un{\lp M \cdot \tilde{n} \rp}{i} - 
	\un{\tilde{n}}{i}\ \lp \tilde{n} \cdot M \cdot \tilde{n} \rp \rc \Big|_{\eta}^{\eta_o} - 
	\int_{ \eta}^{ \eta_o} \dd \eta\ \p{i}^{\perp} \lp \tilde{n} \cdot \tb + 
	\tilde{n} \cdot M \cdot \tilde{n} \rp \,.
	\label{eq:ni-1st-sol}
\end{align}
%
%

\paragraph{Coordinates of the photon's path}

Since $ q^{a} = \dd x^{a}/ \dd \eta$, the coordinates of the photon's trajectory are simply given by 
$ x^{a} = \int \dd \eta\ q^{a}$. There is one important point we want to stress here. Since $ q^{i}$ depends on the fields $ \beta^i$ and $ M_{ij}$, the coordinates of the photon's path will depend on these quantities. That means for instance that under a gauge transformation the coordinates $ x^{i}$ will necessarily change. The same happens if we consider two different realizations of the Universe, each one with the same background evolution but with different field perturbations (that is, different $ \beta^i$ and $ M_{ij}$). On the other hand, by construction $ q^{0} = 1$, so the coordinate $ x^{0} = \eta$ is independent on these fields, and so $ x^{0} = \eta$ is insensible to any gauge transformation.%
\footnote{Though it is sensible to the introduction of new physical field perturbations, or different Universe realizations. This is so, because field perturbations will affect the energy-momentum tensor which determines the time-evolution via the Einstein's equations. Even if the perturbations are small, they give a back reaction on the background \cite{Abramo:1997hu,Mukhanov:1996ak}.%
} %
By construction, the value of $ x^{0} = \eta$ is totally determined by the background evolution of the Universe (or the FLRW spacetime), in particular it is defined by the hypersurface of constant $ \braket{T}$ (see \refs{sec:gSW}, and \cite{Mirbabayi:2014hda}).

To obtain $ x^{i}$, we will use 
\begin{align}
	\int_{\eta_e}^{\eta_o} \dd \eta \int_{\eta}^{\eta_o} \dd \eta' f( \eta') 
	= \int_{\eta_e}^{\eta_o} \dd  \eta \lp \eta - \eta_e \rp f(\eta)\,,
\end{align}
and the relation $ \un{\tilde{n}}{i}_{o} = - \lp \un{ \tb}{i} + q^{i} + M_{ij}\ \un{\tilde{n}}{j} \rp \Big|^{ \eta_o}$ which is valid up to first order. So from \eq{eq:qi-1st-sol} we get 
\begin{align}
	x^{i}  & = x^{i}_{o} +  
	\lc \un{\tilde{n}}{i} - \un{\lp M \cdot \tilde{n} \rp}{i} + 
	\un{\tilde{n}}{i}\ \lp \tilde{n} \cdot M \cdot \tilde{n} \rp \rc \Big|^{\eta_o} \lp \eta_o - \eta \rp
	\label{eq:xi-1st-sol}\\
 	& + \int_{ \eta}^{ \eta_o} \dd \bar{\eta}\ \lc \un{ \tb}{i} + 2 \un{\lp M \cdot \tilde{n} \rp}{i} - 
	\un{\tilde{n}}{i}\ 	\lp \tilde{n} \cdot M \cdot \tilde{n} \rp \rc - 
	\int_{ \eta}^{ \eta_o} \dd \bar{\eta}\ \lp \bar{ \eta} - \eta \rp \p{i}^{\perp} 
	\lp \tilde{n} \cdot \tb + \tilde{n} \cdot M \cdot \tilde{n} \rp \,. 
	\notag
\end{align}
In Eqs.~\eqref{eq:qi-1st-sol}-\eqref{eq:xi-1st-sol} all the integrations are along the unperturbed path. Note that we have parametrized $ q^{i}, \un{\tilde{n}}{i}$ and $ x^{i}$ in terms of the conformal time $ \eta$. This is in contrast with several other works, in which the coordinates $ x^{\mu}$ and the four-momentum $ p^{\mu}$ are obtained in terms of the affine parameter. See for instance Eqs.~(2.20)-(2.24) of \cite{Mollerach:1997up}. For comparison, note that \eq{eq:qi-1st-sol} can be obtained by properly (i.e. by taking into account our \eq{eq:qDEF}) dividing Eqs.~(2.22) by  Eqs.~(2.20) of \cite{Mollerach:1997up}.

We have now all the elements to compute the logarithmic anisotropies up to second order.

\subsection{Sachs-Wolfe at second order: photon's curved path}
\label{sec:SW2nd-curved}

In this section we expand the logarithmic anisotropies up to second order. We will keep quantities evaluated along the photons's curved path. In the next subsection, we express each quantity along the unperturbed path (the Born approximation). Let's start with the ISW. Firstly, it follows from \eq{eq:Adef} that up to second order $ A_{\mu} = \p{\mu} M + [\p{\mu} M,M]/2$, then from \eq{eq:I0} we get
\begin{align}
	I_0  & = \int_{ \eta_e}^{ \eta_o} \dd\eta \lb 
	\tb \cdot \tb' + \tilde{n} \cdot M' \cdot \tb + 
	\lc  \tilde{n} \cdot \tb' + \tilde{n} \cdot \lp M' + 
	\f{[M',M]}{2} \rp \cdot \tilde{n} \rc \lp 1 - \tb \cdot \tilde{n} \rp \rb \notag \\
	& = \int_{ \eta_e}^{ \eta_o} \dd\eta  \lc 
	\lp \tb' + \tilde{n} \cdot M' \rp  \cdot \tilde{n} + 
	\lp \tb' + \tilde{n} \cdot M' \rp  \cdot \tb_\perp \rc \,, 
 	\label{eq:Io2nd}
\end{align}
where $ \tb_\perp$ is the orthogonal projection of $ \tb$ on $ \tilde{n}$, that is, $ \tb_\perp = \tb - \tilde{n} ( \tilde{n} \cdot \tb)$. We have also used the fact that $ \tilde{n} \cdot [M',M] \cdot \tilde{n} = \un{\tilde{n}}{i}\ \lp M_{ik}' M_{kj} - M_{ik} M_{kj}' \rp \un{\tilde{n}}{j} = 0 $.

From \eq{eq:Io2nd} we see that there are two kind of contributions to the ISW. The term that is explicitly linear in the fields is projected along the direction $ \tilde{n}$, while the one which is quadratic in the fields is projected in an orthogonal direction to $ \tilde{n}$.%
\footnote{Of course the term $ \tilde{n} \cdot M'$ is a vector formed by the projection of $ M$ onto $ \tilde{n}$. In that sense, the full term $ \tilde{n} \cdot M' \cdot \tb_\perp $ represents a ``double'' projection of $ M$, one along $ \tilde{n}$ and other along $ \tb_\perp$ which is perpendicular direction to $ \tilde{n}$.
} %
Note also that, $ \lp \tb' + \tilde{n} \cdot M' \rp  \cdot \tb_\perp = \lp \tb' + \tilde{n} \cdot M' \rp_\perp  \cdot \tb = \lp \tb' + \tilde{n} \cdot M' \rp_\perp  \cdot \tb_\perp$, where $ \lp \tb' + \tilde{n} \cdot M' \rp_\perp$ is defined in the same manner as $ \tb_\perp$.

To compute the logarithmic anisotropies, it remains to expand the Doppler effect up to second order. It can be written as
\begin{align}
	\ln \lp \f{\gamma_e \lp 1 - n_e \cdot v_e \rp}{\gamma_o \lp 1 - n_o \cdot v_o \rp} \rp = 
	\lc - v \cdot n + \f{v^{2} - \lp v \cdot n \rp^{2} }{2} \rc \Big|_{ \eta_o}^{\eta_e}  = 
	\lc - v \cdot n + v \cdot \f{v_\perp}{2} \rc \Big|_{ \eta_o}^{\eta_e} \,.
	\label{eq:doppler2ndv1}
\end{align}
Again we see the same behavior as for the ISW effect. That is, terms that are linear in the fields (here $ v$) are projected along%
\footnote{We remind the reader that $ n$ is the direction of observation in the $u $-frame, while $ \tilde{n}$ is the direction of observation in the tetrad-comoving-frame.
} %
 $ n$, while the quadratic terms only receive contribution from the orthogonal direction to $ n$ (here, $ v_\perp$).

Finally, by using \eq{eq:nTOncom} (see also \eq{eq:ToTeExpI0-ini-Eq}) we can also write
\begin{align}
	\ln \lp \f{\gamma_e \lp 1 - n_e \cdot v_e \rp}{\gamma_o \lp 1 - n_o \cdot v_o \rp} \rp   = 
 	 \lc (\beta_{i} - \unI{u}{i})\ \un{\tilde{n}}{i} + \f{ \beta_i \beta^i_\perp - 
 	 \unI{u}{i}\ \un{u}{i}_\perp }{2} \rc \Big|_{ \eta_o}^{\eta_e} \,. 
	%
	\label{eq:doppler2ndv2}
\end{align}
Depending on the situation, one can find it more convenient to use either the first or the second version of the Doppler effect (\eq{eq:doppler2ndv1} or \eq{eq:doppler2ndv2}). We will take the latter, as it involves $ \tilde{n}$. Before going further, we write the previous equation in a covariant manner by using the velocity of comoving observer w.r.t the tetra-frame $ \tb$ and the velocity of the observer $ u$ w.r.t the tetrad-frame, that is, $  \un{ \vf }{a} \equiv \lp 0, \un{u}{i}/ \un{u}{0} \rp$. Up to second order we have 
\begin{align}
	\un{ \tb}{a} = \lp 0, \beta^{i}\rp \qquad\text{and}\qquad  \un{ \vf }{a} = \lp 0, \un{u}{i} \rp \,.
\end{align}
Here, the subscript $ F$ is because we can think of the observer $ u$ as being a fish moving through a river (the tetrad frame). This idea is explored in a companion paper \cite{Roldan:2017kgd}. Joining all the previous results we have up to second order
\begin{align}
	\Theta  = \lp \cT + \Phi + (\tb - \vf) \cdot \tilde{n}  \rp  & \Big|_{ \eta_o}^{\eta_e} + 
	\int_{ \eta_e}^{ \eta_o} \dd\eta  \lp \tilde{n} \cdot \tb' + 
	\tilde{n} \cdot M' \cdot \tilde{n} \rp \notag \\
	& \qquad  + \f{ \tb \cdot \tb_\perp - \vf \cdot \vf^{ \perp} }{2}\Big|_{ \eta_o}^{\eta_e} + 
	\int_{ \eta_e}^{ \eta_o} \dd\eta  \lp \tb' + 
	\tilde{n} \cdot M' \rp  \cdot \tb_\perp \,.
	\label{eq:sw2nd-curved}
\end{align}
We have written the logarithmic anisotropies in this way to stress that the first line is formally%
\footnote{I said formally, because here each quantity is considered up to second order.
Additionally they evaluated along the photon's curved path, while at first order, $ \Theta$ is computed using the background trajectory.
} %
equal to the first-order logarithmic anisotropies. So (formally), the difference comes only from the second part. These two lines are different in nature, so they could be measured independently. Note that $ - \tilde{n}$ is the direction of propagation of photons as seem by the tetrad-comoving-observers, so the plane perpendicular to $ \tilde{n}$ is the plane of the photon's polarization. We conclude that only the projection on the plane of polarization of the field perturbations, contribute to the explicitly quadratic terms of $ \Theta$ (second line of \eq{eq:sw2nd-curved}).

\subsection{Sachs-Wolfe at second order: the Born approximation}
\label{sec:SW2nd-Born}

In this section the second-order logarithmic anisotropies are given by using the Born approximation, that is, we express each quantity along the path $ x^{i} = x^{i}_{o} + \un{\tilde{n}}{i}_{o} \lp \eta_o - \eta \rp$, which is the path inferred by the observer ignoring perturbations. This can be useful for numeric computations and also because the notation of previous results in literature (e.g., those in \cite{Roldan:2016ayx}) is closer to the one we use below.

We will define the deviation $ \dx^{i}$ from the Born approximation by the relation 
$ x^{i} = \lc x^{i}_{o} +  \un{\tilde{n}}{i}_o \lp \eta_o - \eta \rp \rc + \dx^{i} $. Analogously, we will write $ \un{\tilde{n}}{i} = \un{\tilde{n}}{i}_o + \un{\dn}{i}$. The explicit expression of $ \dx^{i}$ and $ \un{\dn}{i}$, follow directly from Eqs.~\eqref{eq:ni-1st-sol} and \eqref{eq:xi-1st-sol}. With theses definitions we can Taylor expand the logarithmic temperature anisotropies around the Born approximation's path, as
\begin{align}
	\Theta & = \Theta_{Born} + \lc \delta \cT + \delta \Phi + \tilde{n} \cdot \delta (\tb - \vf) + 
	 (\tb - \vf) \cdot \dn \rc_{e} \notag \\
	& + \int_{ \eta_e}^{ \eta_o} \dd\eta\ \lc \lp \tilde{n} \cdot \delta \tb' + 
	\tilde{n} \cdot \delta M' \cdot \tilde{n} \rp  + \lp \dn \cdot \tb' + 
	2 \dn \cdot M' \cdot \tilde{n} \rp \rc \,, 
	\label{eq:sw2nd-born}
\end{align}
where $ \Theta_{Born}$ is the same as \eq{eq:sw2nd-curved} but with everything evaluated in the Born approximation. Here the notation must be intuitive. For instance, $ \lp \delta M' \rp_{ij} = \dx^{k} \p{k} M_{ij}' $ with similar expressions for the other fields. The only difference is with the intrinsic logarithmic anisotropies $ \cT_{e}$ which in general will depend not only on the position $x_e$ but also on the direction of emission $ - \tilde{n}_{e}$, that is, $ \cT_{e} = \cT (x_e, - \tilde{n}_{e})$. So we must use
\begin{align}
	\delta \cT_e = \lp \dx^{i}_e \p{i}  + \un{\dn}{i}_e \pp{}{\un{\tilde{n}}{i}_e} \rp \cT_e \,.
\end{align}
%
%

\subsection{Lensing and time-delay}
\label{sec:lensing}


To end this section, we remind the reader about the concepts of lensing and time-delay, which are encoded into $ \dx^{i}$ and are correlated with the ISW. To obtain the time-delay, we project $ \dx^{i}$ along the radial direction (see \eq{eq:xi-1st-sol})
\begin{align}
	\dx^{i} \unI{\tilde{n}}{i} = 
	\int_{ \eta}^{ \eta_o} \dd \bar{\eta}
	\lp \tilde{n} \cdot \tb + \tilde{n} \cdot M \cdot \tilde{n} \rp \,,
	\label{eq:TDelay}
\end{align}
this quantity tells us that photons are not coming from a spherical shell of radius $ r$ 
but from a distorted surface whose ``radius'' in direction $ \tilde{n}$ is distorted by $(\dx^{i} \unI{\tilde{n}}{i})_{e}$. There are two types of lensing terms: the first one is given by the transverse component of $ \dx^{i}$,
\begin{align}
	\dx^{i}_\perp = - \un{\lp M_o \cdot \tilde{n}_o \rp}{i}_\perp \lp \eta_o - \eta \rp + 
	\int_{ \eta}^{ \eta_o} \dd \bar{\eta}\ \lc \un{\tb}{i}_\perp + 2 \un{\lp M \cdot \tilde{n} \rp}{i}_\perp \rc - 
	\int_{ \eta}^{ \eta_o} \dd \bar{\eta}\ \lp \bar{ \eta} - \eta \rp \p{i}^{\perp} 
	\lp \tilde{n} \cdot \tb + \tilde{n} \cdot M \cdot \tilde{n} \rp \,,
	\notag
\end{align}
and the second one is just the local deflection angle $ \dn$, which from \eq{eq:ni-1st-sol} is 
\begin{align}
	\un{\dn}{i} = - \lc \un{\lp M \cdot \tilde{n} \rp}{i}_\perp\rc \Big|_{\eta}^{\eta_o} - 
	\int_{ \eta}^{ \eta_o} \dd \eta\ \p{i}^{\perp} \lp \tilde{n} \cdot \tb + 
	\tilde{n} \cdot M \cdot \tilde{n} \rp \,.
	\label{eq:deflection}
\end{align}
Lensing and time-delay are correlated with the ISW effect due to the second line of \eq{eq:sw2nd-born}.
Note that, regarding the logarithmic anisotropies $ \Theta$, these are the only quantities that are correlated with the ISW. This is not true however for $ \Delta T_o/\bar{T}_o$ which involves powers of $ \Theta$ and therefore will automatically correlate the ISW with other terms like $ \cT$ and $ \Phi$. Because of that, making maps of the logarithmic anisotropies will provide an optimal tool for study the ISW.

Below we briefly comment on the comparison with other works and also briefly cite some of the results that will publish in a companion paper.

\section{Future work and conclusions}

\subsection{Future work and comparison with literature}

The results of appendix A of \cite{Roldan:2016ayx} (and therefore, the results of \cite{Mollerach:1997up}) are equivalent to the ones given in the previous section.%
\footnote{Although in \cite{Roldan:2016ayx} we took the perturbations to vanish at the observer position.
} %
In particular compare \eq{eq:sw2nd-born} with Eqs.~A.32-A.35 of \cite{Roldan:2016ayx}. In comparing the results of \cite{Roldan:2016ayx}, we must take into account the following relationship
\begin{align*}
	\un{\de}{i} = \un{\dn}{i} + \un{\lp \tb + \tilde{n} \cdot M \rp}{i}_{\perp} \,,
\end{align*}
where the quantity $ \de$ was defined in Eq.A.29 of \cite{Roldan:2016ayx}. The previous relation follows directly from \eq{eq:deflection} and Eq.A.28 of \cite{Roldan:2016ayx}. This shows that the interpretation given in \cite{Roldan:2016ayx} for the quantity $ \de$ as the \textit{local deflection angle} is wrong, because the true local deflection angle is given by $ \un{\dn}{i}$. Apart from this fact, the results of \cite{Roldan:2016ayx} are correct. A more detailed comparison of our results with those already present in literature will be discussed in a future paper \cite{Roldan:2017-2}.

On the other hand, since we have introduced several new concepts: a new parametrization of the metric, the logarithmic intrinsic temperature anisotropies $ \cT$, the direction of observation by tetrad-comoving-observers $ \tilde{n}$, etc, the gauge transformations of theses quantities have not been discussed before in literature. In addition, gauge transformations when applied to the CMB anisotropies involves several subtle issues as it was firstly discussed in \cite{Mirbabayi:2014hda}. In a companion paper \cite{Roldan:2017kgd}, we will discuss the gauge transformations of the relevant quantities introduced in this paper and explicitly show the gauge invariance of our second-order formula \eq{eq:sw2nd-born}. Special emphasis is put on the subtle issues of gauge transformations on the CMB. 


\subsection{Conclusions}

We have obtained the non-linear generalization of the Sachs-Wolfe + integrated Sachs-Wolfe formula describing the CMB temperature anisotropies \eq{eq:Gsw}. Our result is valid at all orders in perturbation theory, includes scalar, vector and tensor perturbations, and is valid in any gauge. Direct observational consequences of our result have been discussed, in particular the fact that the logarithmic temperature anisotropies $ \Theta = \ln \lp 1 + \Delta T_o/ \bar{T}_o \rp $ is more suitable for data analysis than the usual temperature anisotropies $ \Delta T_o/ \bar{T}_o$. The reason is that by taking the logarithm we automatically remove many secondary effects which otherwise would bias the analysis of the data. This will be of particular importance for the search of primordial non-Gaussianity and for analysis of the ISW effect and lensing. \\
\noindent
Then we expanded our exact expression up to second order and got results which are very simple and intuitive, see Eqs.~\eqref{eq:sw2nd-curved} and \eqref{eq:sw2nd-born} for two different versions. Finally, several concepts have been introduced as the logarithmic intrinsic anisotropies $ \cT$ and $ \cT_o$ (see \refs{sec:gSW}), the tetrad-comoving-observers in \refs{sec:Photons-and-obs} and an useful parametrization of the metric \refs{sec:metric-tetrad} which expresses the $ 0 - i$ metric components in terms of the four-velocity of comoving observers.

\section*{Acknowledgments}

I thank Thiago Pereira and Elvis Soares for useful discussions and suggestions. I also thank Mauricio Calv\~ao for introducing me to the 1 + 3 - covariant formalism, and for useful discussions on aberration. I thank Cyril Pitrou for drawing my attention on the importance of the logarithmic transform on spectral distortions. Finally, I thank the anonymous referee of paper \cite{Roldan:2016ayx} because his (her) useful comments somehow influenced the style of this article.

 \appendix

\bibliographystyle{JHEP2015}
\bibliography{refs-up-to-1999,refs-2000-2009,refs-2010-2019}

\end{document}